\documentclass[aps,prc,preprint,superscriptaddress]{revtex4}

\usepackage{graphicx}

\begin{document}
\title{Evidence for  higher nodal band states with   $^3$He cluster structure
 in $^{19}$Ne
 and prerainbows in $^3$He+$^{16}$O scattering }
\author{S. Ohkubo$^1$   and   Y. Hirabayashi$^2$ }

\affiliation{$^1$Department of Applied Science and Environment,
Kochi Women's University, Kochi 780-8515, Japan  }

\affiliation{$^2$Information Initiative Center,
Hokkaido University, Sapporo 060-0811, Japan}

\date{\today}

\begin{abstract}
\par
 The existence of a    higher nodal band state with  a $^{3}$He
 cluster structure, i.e.  
a vibrational mode in which 
 the inter-cluster   relative motion is excited, 
  in  $^{19}$Ne in addition to those with the $\alpha$ cluster structure 
in $^{20}$Ne and the $^{16}$O cluster structure in $^{32}$S, is suggested,
which     reinforces 
the  importance of the concept of $^3$He-clustering in nuclei.
This conclusion was reached  by investigating  
  $^{3}$He scattering from $^{16}$O in a wide range of incident energies 
and   prerainbow oscillations.  
 \end{abstract}


\maketitle


\par
 \par 
The existence of an excitation mode
 of  inter-cluster   relative motion  is  essentially  characteristic to the cluster
structure in nuclei like   
 a phonon excitation mode of the  vibrational motion  in spherical nuclei.
 The purpose of this paper is to show for the first time  that this  excitation,
 i.e. higher nodal
 states, exists for a $^3$He cluster  by
 studying  $^3$He scattering from $^{16}$O which shows  prerainbow oscillations.

\par
  The $\alpha$ cluster structure is widely understood in light nuclei and 
 in the typical  heavy nuclei.
 In contrast to an atomic  molecule like the  hydrogen molecule where many clear
inter-atomic  vibrational
excitations are observed, the  vibrational excitation of the inter-cluster  
 relative motion in  nuclei is rarely observed because of the weak attractive
potential between the constituent clusters. The $K=0^+_4$ band
 starting  at 8.6 MeV
in $^{20}$Ne is an example observed in light nuclei.

 On the contrary to a naive picture that the cluster structure is broken in heavier
 nuclei
because of a strong spin-orbit potential, it is  known \cite{Michel1998,Yamaya1998}
 that the  $\alpha$ cluster structure
 persists typically in $^{44}$Ti in the fp-
shell region.  
 The prediction and observation  of the higher nodal bands with the $\alpha$+$^{40}$Ca cluster
 structure in $^{44}$Ti and  the $\alpha$+$^{36}$Ar cluster structure in $^{40}$Ca, 
respectively 
\cite{Michel1998,Yamaya1998}
gave further  foundation to the $\alpha$ cluster picture  in heavier nuclei.

 \par In addition to the $\alpha$ particle,  $^{16}$O  is a tightly-bound 
 doubly-magic nucleus. The famous gross structure observed in the 90$^\circ$ excitation function 
in $^{16}$O+$^{16}$O elastic scattering  has been discussed for many years in relation 
to the $^{16}$O+$^{16}$O  cluster structure in $^{32}$S \cite{Gobbi1979}, 
which is an analogue of $^8$Be with 
 $\alpha$+ $\alpha$ structure. Different from $^8$Be and $^{20}$Ne, because of the lack 
of the evidence of a clear experimental rotational  band with the $^{16}$O+$^{16}$O  cluster structure
 near the threshold, whole  $^{16}$O  cluster aspects in
$^{32}$S had  not  been clear. However,  recently  a unified description of  
nuclear rainbows,  prerainbows and the  $^{16}$O+$^{16}$O cluster structure at a 
low excitation 
energy region suggested \cite{Ohkubo2002} that there exists a lowest $N=24$ 
($N=2n+L$ with $n$ and $L$ being the number of the nodes in the relative wave
 function and 
the orbital motion angular momentum, respectively) rotational band allowed  by
 the Pauli principle and that this corresponds well to the observed band states. 
The famous gross resonant structures were also found to be the higher nodal
 states with the $^{16}$O+$^{16}$O  configuration with $N=28$, in which
 the inter-cluster 
relative motion is excited by two more nodes compared with the lowest $N=24$ 
 band, which 
corresponds to the superdeformed structure in $^{32}$S.

  Thus the concept of an  excitation mode  of the inter-cluster  relative 
 motion 
 with a cluster structure  has been  established for the 
$\alpha$ cluster  and $^{16}$O cluster theoretically and  experimentally.

\par
 It is  known    that the  A=3 cluster, i.e. triton and $^3$He cluster, is
 also important
for   understanding  the structure of  nuclei such as $^7$Li, $^7$Be and $^{19}$F. 
To reinforce the validity of the  A=3 cluster it is  important to know if 
 the concept of a  higher nodal state is universal and if the  excitation of 
the inter-cluster   relative motion   exists also for the   $^3$He 
 and t clusters. 
 However,  to the authors' best knowledge,
the existence of such a higher nodal state  has not been confirmed
 experimentally and theoretically. 
 In fact, Waltham {\it et al.} \cite{Waltham1983}  experimentally studied the $^3$He 
cluster structure in $^{19}$Ne,
 paying particular attention to the  possible
existence of  higher nodal states with  
$^3$He +$^{16}$O structure and unfortunately  reached a negative  conclusion.
 In contrast to $^{19}$Ne, the triton cluster structure of the mirror nucleus
$^{19}$F was studied  theoretically  by Buck and Pilt \cite{Buck1977} 
with a cosh potential 
and by Sakuda {\it et al.} \cite{Sakuda1979A} in a semi-microscopic model
 and experimentally  by many authors    \cite{Ajzenberg-Selove1987,Hamm1976}.  
 However,  a  higher nodal state with the
 t+$^{16}$O configuration has not been identified.

\par
To reveal  the cluster structure of nuclei, it is very useful to 
study not only the low-lying bound and quasi-bound states of the composite system
 but also the  scattering 
in a  unified way because this checks the interaction
 potential 
 not only in the surface region but also in the internal region.
 In fact, in this approach
 a long-standing controversy about the existence of the cluster structure in 
  $^{44}$Ti   was successfully solved \cite{Michel1998,Michel1986,Ohkubo1988}.
As for the $\alpha$ cluster structure in nuclei,
 an $\alpha$-nucleus potential has been studied by a systematic analysis of $\alpha$ particle 
scattering from nuclei in a wide range of incident energies. For example, for the 
 typical nuclei
like $^{16}$O and $^{40}$Ca, a unique global optical potential has been established 
\cite{Michel1998,Michel1983,Delbar1978}. These systems have been a prototype for the study of the interaction 
potential 
of the composite particles and the $\alpha$ cluster structure of the compound system.
 The global
  real potential  is described well   by a phenomenological 
potential with a form factor of Woods-Saxon squared  or a folding model   rather than
a conventional  Woods-Saxon  potential 
\cite{Ohkubo1977,Ohkubo1988,Abele1993,Atzrott1996,Michel1983,Delbar1978}.  This
 potential is  powerful  in the unified description of bound and scattering
 states of the composite system in the sd-shell,  fp-shell  and  
  much heavier
 regions like $^{94}$Mo  and $^{212}$Po  
 \cite{Atzrott1996,Ohkubo1995}.

\par
 On the other hand, for $^3$He a unique global potential has not been established  and
 a  unified description of bound  states and scattering has scarcely been  achieved.
  The spin-orbit potential is  an  interesting 
  and challenging subject
 for this system and by using a polarized beam  an extensive study was done 
by the Birmingham
 group \cite{Lui1980}. 
To understand 
the $^3$He-nucleus interaction it is important to establish  a unique global
  central real potential first.
There are systematic experimental data of angular distributions in 
 $^3$He scattering from $^{16}$O at  
$E_L$=15 to 60 MeV
\cite{Lui1980,Vernotte1982,Alvarez1982,Trost1987,Adodin1992}.
  Analyses of $^3$He scattering  have been  done mostly
 by using a conventional Woods-Saxon potential.
  A folding model was   applied to $^3$He+$^{16}$O scattering by
Khallaf {\it et al} \cite{Khallaf1997}.
 However, there has been no systematic double folding model
 analysis 
of the  $^3$He+$^{16}$O system  from the viewpoint of a unified description of
 bound 
and scattering states.

The double folding potential we use is given as follows:
\begin{equation}
V_{ij}({\bf R}) =
\int \rho^{\rm (^3He)} ({\bf r}_{1})\;
     \rho^{\rm (^{16}O)} ({\bf r}_{2})\;
v_{\rm NN} (E,\rho,{\bf r}_{1} + {\bf R} - {\bf r}_{2})\;
{\rm d}{\bf r}_{1} {\rm d}{\bf r}_{2}  \; ,
\end{equation}

\noindent
where $\rho^{\rm{ (^3He)}} ({\bf r})$ is the ground 
state density
of  $^3$He taken from Cook {\it et al.} \cite{Cook1981} and 
$\rho^{\rm (^{16}O)} ({\bf r})$ is  the  nucleon density of  
$^{16}$O, which is obtained from 
the charge-density distribution determined by electron scattering \cite{deVries1987}
 after the deconvolution of the proton size in the  usual way, while $v_{\rm NN}$ denotes
 the DDM3Y interaction \cite{Kobos1984}.
In the analysis   we  introduce a  normalization factor 
 $\lambda$ for  the real part of the potential and phenomenological
 imaginary potentials with a  Woods-Saxon form factor.

\begin{figure}[htb]
\includegraphics[width=8cm]{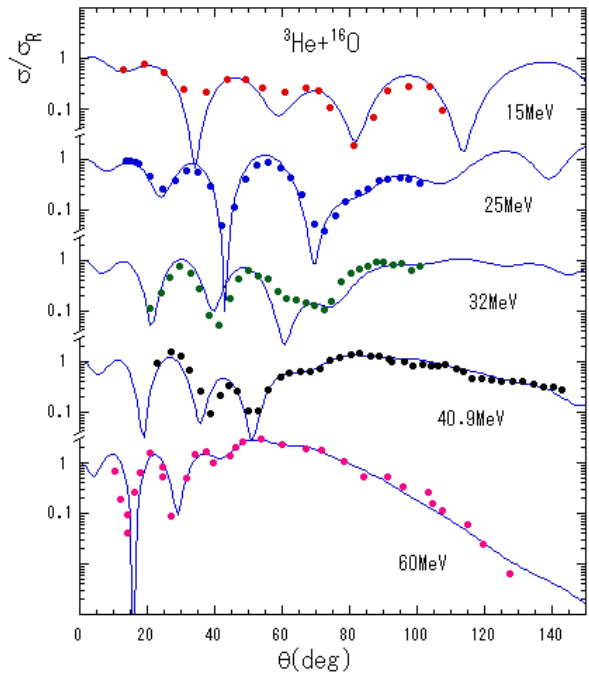}
 \protect\caption{\label{fig.1} {(Color online)
Comparison of the 
double folding model cross sections (solid lines)  with the experimental  
 angular distributions  (points)  
\cite{Lui1980,Vernotte1982,Alvarez1982,Trost1987,Adodin1992}.
 }
}
\end{figure}

In  Fig.~1   calculated angular distributions  of $^3$He+$^{16}$O scattering  are displayed 
in comparison  with 
 the   experimental data. 
In the calculations the renormalization factor  $\lambda$ and
imaginary potential parameters are adjusted to fit the experimental  data.
The  imaginary potential parameters and  the properties of the real potentials used are given 
in Table I. 
The energy evolution of the characteristic angular distributions 
is  well reproduced.
At $E_L$=25 MeV  a large radius parameter of the imaginary potential is needed to 
fit the  angular distribution beyond 70$^\circ$. This is consistent with the 
analysis by   Vernotte  {\it et al.} \cite{Vernotte1982}. 
 The calculated angular distributions
show some remnant  of   Anomalous Large Angle Scattering (ALAS) or
 Backward Angle Anomaly (BAA),
although unfortunately there are no data at extreme backward angles.
The rise at the backward angles is not so pronounced compared with  typical 
$\alpha$+$^{16}$O scattering at the corresponding energies. 
At $E_L$=32 MeV the angular distribution shows  a prerainbow oscillation, which appears 
in the transitional energies  from ALAS-like  behavior 
to rainbow scattering. At $E_L$=40.9 MeV  a  prerainbow oscillation is also seen  
 with a deep minimum at $\theta$$\approx$50$^\circ$ and  a plateau beyond.
 The typical  fall-off
 of the angular distribution, which corresponds to the dark side of the rainbow,
 appears at $E_L$=60 MeV with the  first order  Airy minimum
 at  $\theta$$\approx$27$^\circ$.

\begin{table}[th]
\begin{center}
\protect\caption{ The normalization factor $\lambda$, volume integral per nucleon
pair $J_V$ of the folding potential,  parameters of the imaginary potentials
and its volume integral per nucleon
pair $J_W$
for the $^3$He+$^{16}$O system in the conventional notation. 
 The calculated rms radius  $\sqrt{<R^2>}$ of the real potential is 3.74 fm for 
all the  incident energies. }
\begin{tabular}{ccccccccc}
 \hline
  \hline
  $E_L$ & $\lambda$  &      $J_V$      & $W$ & $R_I$ & $a_I$ & $J_W$ \\
  (MeV) &         & (MeVfm$^3$) & (MeV)& (fm) & (fm) &  (MeVfm$^3$)  \\
 \hline
 15   & 1.24    &    431.6    &  16 & 4.4 & 0.10  &  119.5 \\ 
 25   & 1.23    &    417.7    &  6  & 5.8 & 0.45  &  108.2 \\ 
 32   & 1.24    &    412.9    & 12  & 4.4 & 0.80  &  118.3  \\ 
 40.9 & 1.24    &    408.2    & 12  & 4.5 & 0.90  &  133.1    \\ 
 60   & 1.24    &    388.8    &  12 & 4.6 & 0.80  &  132.4  \\ 
 \hline
 \hline
\end{tabular}
\end{center}
\end{table}
\begin{figure}[htb]
\includegraphics[width=8cm]{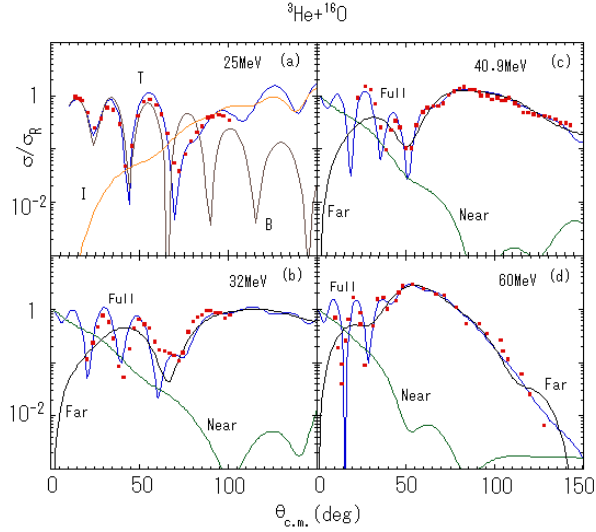}
 \protect\caption{\label{fig.2} {(Color online) 
Double folding model cross sections  (solid lines) 
are  decomposed into the internal-wave (dashed lines)  and the barrier-wave (dotted lines) 
contributions at 25 MeV (a),  and  the  farside (dashed lines) and the  nearside (dotted lines) 
 contributions at 32, 40.9 and 60 MeV (b)-(d), and compared with the experimental
 data \cite{Lui1980,Vernotte1982,Alvarez1982,Trost1987,Adodin1992}.
}
}
\end{figure}

\par
To see the evolution of the Airy minimum the  calculated angular distributions decomposed 
into farside and nearside contributions following the Fuller's subscription \cite{Fuller1975}  are 
displayed in Fig.~2 (b-d) at $E_L$=32, 40.9 and 60 MeV.   The Airy minimum is clearly seen
 in each angular distribution, which shows that   $^3$He+$^{16}$O scattering  is 
  transparent.  This transparency can be further seen in Fig.~2(a) by decomposing the 
scattering amplitude into the  internal-wave subamplitude, which penetrates deep into the internal 
region of the potential  and the barrier-wave subamplitude, which is reflected at the 
barrier \cite{Brink1985,Albinski1982}. 
Although ALAS, which is typically observed in 
  $\alpha$+$^{16}$O scattering \cite{Ohkubo1977,Michel1983}, is not clearly seen 
 in $^3$He+$^{16}$O scattering due to a lack of the experimental data at backward 
angles,
 there is an enhancement of cross sections at large angles  in the calculated 
angular distributions
 due to the internal-wave contributions.
The sharp minimum at  70$^\circ$ is due to the interference between the internal
waves
 and the  barrier waves and is a prototype of the  Airy minimum of the 
prerainbow  at 32 MeV and the rainbow at 60 MeV   in Fig.~2(b)-(d).

 \begin{figure}[tbh]
\includegraphics[width=8cm]{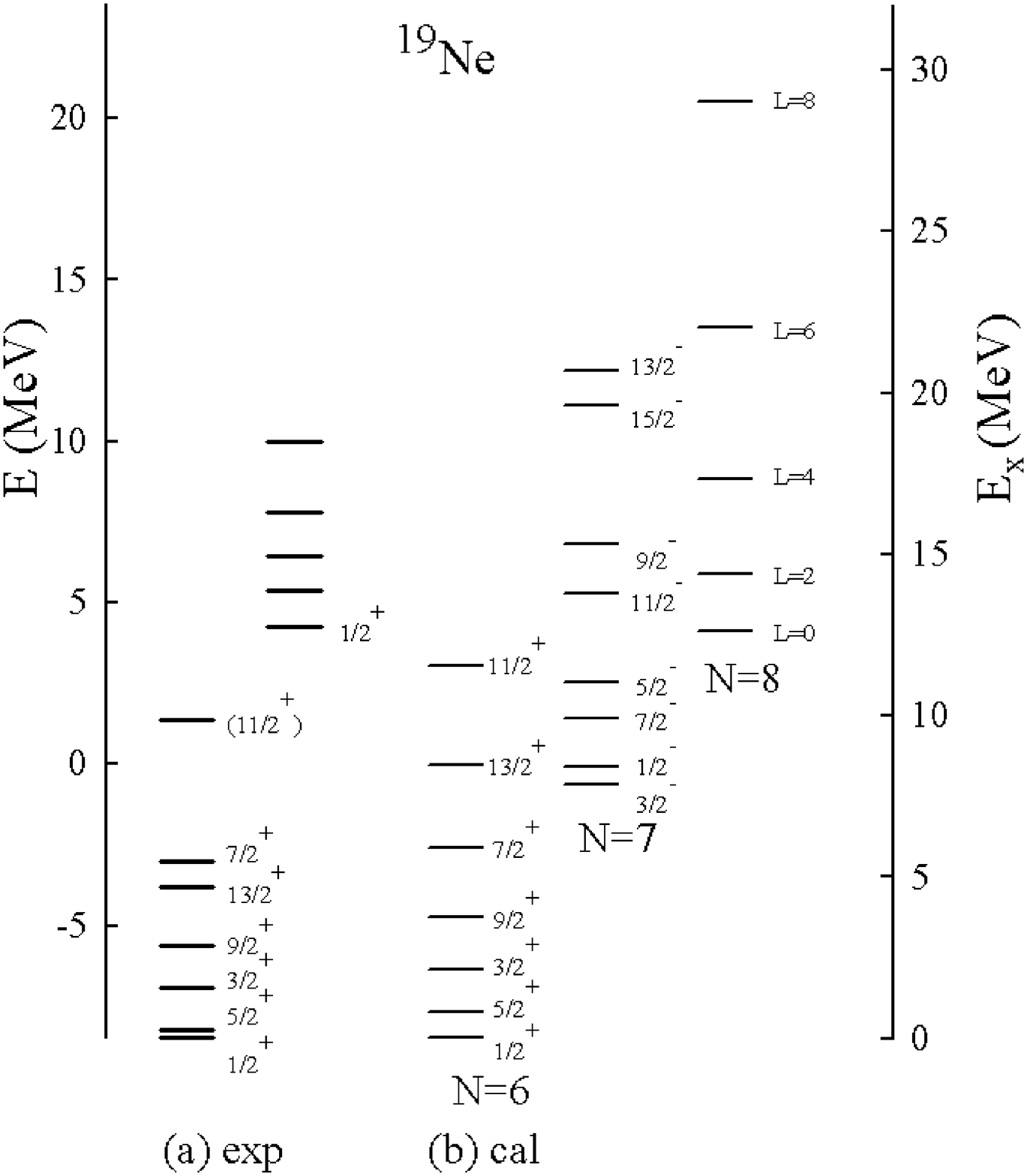}
 \protect\caption{\label{fig.3} {(a) The experimental  
 $^3$He cluster  
 state candidates in  $^{19}$Ne \cite{Hamm1976,Ajzenberg-Selove1987,Waltham1983,Kraus1988}
; (b) The $N=6$, $N=7$ and  $N=8$  
 states supported by the double folding potential  (the potential is fixed to 
 the folding potential at $E_L$=15 MeV with  $\lambda$=1.34). 
Energies are given with respect to the $^3$He+$^{16}$O threshold (left) and in 
excitation energy (right).
  For the $N=8$  band the energies where the phase shift passes 270$^\circ$ for $L=4$, 6 and 8,
and  the energies where the phase shift approaches the  highest for $L=0$ and 2,
   are  displayed.
}
}
 \end{figure}

The obtained real potential   should 
work at the bound and quasi-bound  energy region as was  demonstrated in the typical 
 $\alpha$+$^{16}$O and $\alpha$+$^{40}$Ca systems \cite{Michel1998}.
  The potential determined
at $E_L$=15 MeV locates the lowest Pauli-allowed ground state of $^{19}$Ne
with the $^3$He+$^{16}$O configuration  at -4.97 MeV from the $^3$He threshold, 
which  falls well within the  range of
the  experimental energy -8.44 MeV. 
As seen in Table I, the volume integral of the folding  potential
 shows a tendency to increase as 
the energy  decreases, which arises from the energy dependence of the DDM3Y interaction. 
 To reproduce  the experimental ground state energy 
  $\lambda$=1.34 is used ($J_V$=466.4 MeVfm$^3$) and the calculated energy
 levels are displayed in Fig.~3. 
The  double folding potential used is shown in Fig.~4.
 In the calculation the  following spin-orbit potential
 is introduced:
 \begin{equation}
V_{so}(R) =
- V_{so}{\Bigl(}\frac{\hbar}{m_\pi c}{\Bigr)}^2 \frac{1}{R}  \frac{dV(R)}{dR} 
\vec{L} \cdot \vec{\sigma} ,
\end{equation}
where $\vec{\sigma}$ is the spin of the $^3$He cluster.  The strength constant
 $V_{so}$ = 0.011 MeV is used 
to fit the splitting of the $\frac{5}{2}^+$ ($E_x$=0.238 MeV) and $\frac{3}{2}^+$ (1.576 MeV) 
states.  
The calculated ground band with   $N=6$  corresponds well to the experimental levels.
 The ground band has a rather 
 shell-like structure and   $^3$He clustering  is not 
strong. The ($\frac{11}{2}^+$)  state at $E_x$=9.8 MeV  was observed in $^3$He-transfer reactions
\cite{Kraus1988}.
  The $^3$He cluster strength of the experimental high spin states $\frac{11}{2}^+$ 
 and $\frac{13}{2}^+$   
 of the ground band  could be shared over   two or more states 
  as in $^{19}$F  \cite{Buck1977} and  the theoretical $\frac{11}{2}^+$  and
 $\frac{13}{2}^+$  states in Fig.~3
 could be  the centroid of them.
 As a parity doublet partner of the ground band, the  $N=7$ negative parity band
with the $^3$He+$^{16}$O cluster structure whose band head $\frac{3}{2}^-$ state starts
just  near  the $^3$He threshold is   predicted.  
In  the mirror nucleus $^{19}$F, some of the member states of the  N=7 band with the 
t+$^{16}$O cluster structure  have been 
identified in the cluster model calculations  by Buck and Pilt 
\cite{Buck1977} and Sakuda {\it et al.} \cite{Sakuda1979A}.
 The  ($\frac{7}{2}^-$) state in $^{19}$Ne at
6.861 MeV, which is an isospin-analog state of the $N=7$, $\frac{7}{2}^-$ state 
at 6.927 MeV 
in $^{19}$F \cite{Buck1977,Sakuda1979A,Utku1998} could be a member state of the
 $N=7$ band in $^{19}$Ne.  Sakuda {\it et al.} \cite{Sakuda1979A} 
pointed out that to reproduce 
the experimental energy levels of  the N=7  negative parity band  correctly the
 coupling between the cluster states with
 the t+$^{16}$O  configuration and the cluster states with the $\alpha$+$^{15}$N$^*$
 configuration 
is important, which will also hold   in $^{19}$Ne. 
Very recently Yamazaki {\it et al.} \cite{Yamazaki2007} claim  that they observed  three low-lying 
 members of the negative parity rotational band in the $^{16}$O($^6$Li,t) transfer
 reactions.
 Although no details are given  in Ref.\cite{Yamazaki2007}, 
it could be that these  correspond  to member states of the $N=7$ parity
 doublet band. 

\begin{figure}[tbh]
\includegraphics[width=6cm]{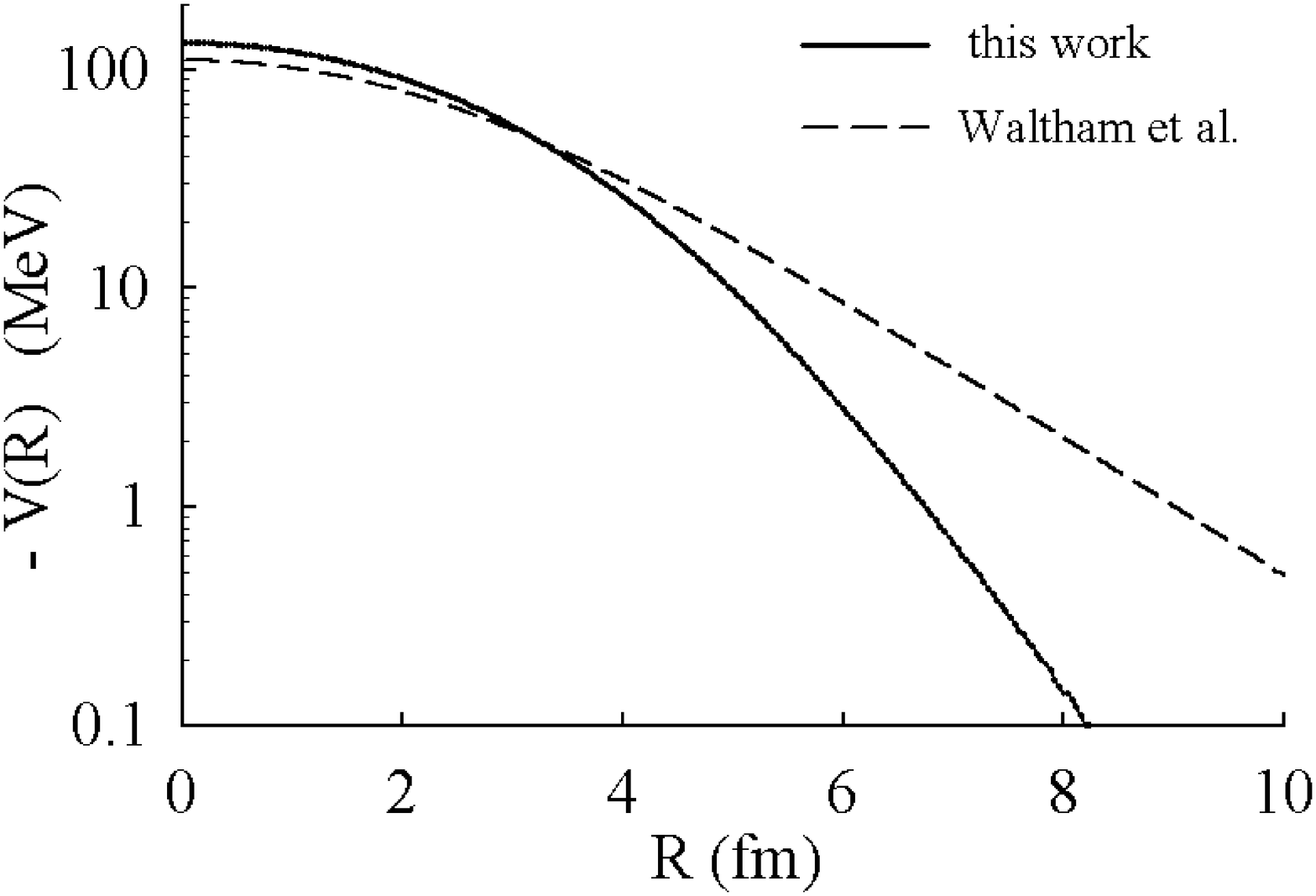}
 \protect\caption{\label{fig.4} {  
Comparison of our central double folding potential used in Fig.~3 (solid line) with 
the cosh potential 
of Waltham {\it et al}  \cite{Waltham1983} (dashed line).
 }
}
\end{figure}

\begin{figure}[tbh]
\includegraphics[width=7cm]{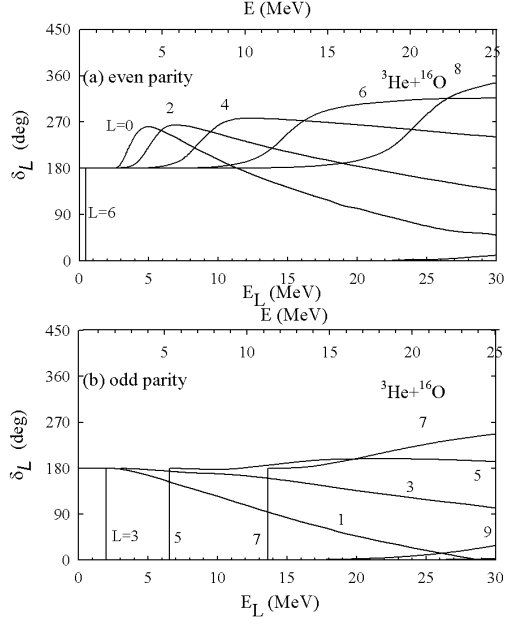}
 \protect\caption{\label{fig.5} {  
Phase  shifts    calculated  with the   double 
 folding potential used in the energy level calculations  in Fig.~3.
 }
}
\end{figure}

 In order to investigate  the cluster states above the threshold 
  phase shifts calculated with this potential  (by switching off the spin-orbit potential) 
are displayed in Fig.~5.  It is noticed that the phase
 shifts for even parity and odd parity show very different behavior.
The phase shifts for even parity increase toward 270$^\circ$  and  the
 $L$=4, 6 and 8 show  broad resonant behavior.  
On the other hand, for odd parity there appears no resonant
 behavior to  approach  270$^\circ$.  This different behavior of the even and odd parity
 phase shifts  is very similar to the case for the $\alpha$+$^{16}$O system.

 In $^{20}$Ne the corresponding 
 states with $N=10$ with a  developed $\alpha$+$^{16}$O cluster structure 
   have one more
node compared with the $N=8$ ground band  wave functions. It was pointed out   
 \cite{Ohkubo1977} that although the high spin  $6^+$ and $8^+$ states
 of the higher nodal $N=10$ band  are difficult to
 be observed as an individual energy level because of its  large width, its persistent 
existence  can be seen
 in the phenomenon of  BAA or ALAS in $\alpha$+$^{16}$O scattering. 
 The  $N=8$ band with the $^3$He+$^{16}$O cluster structure
 in $^{19}$Ne is  an analogue of  the higher nodal band in $^{20}$Ne.  
 As seen in Fig.~2(a), the sharp Airy minimum     of the prerainbow oscillation at 70$^\circ$
in the angular distribution at $E_L$=25 MeV, 
 which clearly separates  the two  components of the amplitude responsible
 for the Airy structure, is caused by the interference between  the internal waves, 
  which are  mostly due  to the existence of the high spin members of the higher nodal band
 and the barrier waves.

 Based on  the above picture,  the  experimental results
 of Waltham {\it et al.}  \cite{Waltham1983} in  $^{16}$O($^3$He,$\gamma$) capture 
 reactions  
 and the  old $^3$He  elastic scattering data at $E_L$=5.05 MeV by R\"{o}pke
 {\it et al.} \cite{Roepke1967}  can now be
 understood quite naturally as  the experimental evidence for  the existence 
 of the $N=8$ higher nodal states.
Firstly the   $\frac{1}{2}^+$ state at $E_x$=12.69 MeV 
  observed in the measurement of $^3$He+$^{16}$O elastic scattering
 at $E_L$=5.05 MeV ($E$=4.25 MeV)
 \cite{Roepke1967,Ajzenberg-Selove1987}
 well corresponds
 to the  calculated higher nodal state with $L=$0 (Fig.~3). 
This state has a large  $^3$He decay 
width $\Gamma_{c.m.}$/$\Gamma_{tot}$=0.43  with $\Gamma_{c.m.}$=0.18 MeV 
 \cite{Roepke1967}    in accordance with the 
character of the higher nodal member state. 
Secondly as for the experimental results of 
Waltham {\it et al.}  \cite{Waltham1983}, they  performed  an  experiment to search for 
 highly excited 
 $^3$He cluster states in $^{19}$Ne and discussed  the newly observed 
energy levels in comparison with the cluster model calculations with a cosh potential of 
Buck and Pilt \cite{Buck1977}. They  reached the   conclusion   that the
 cluster model is unsuccessful in the highly excited  energy region \cite{Waltham1983}.
However, we   note  that the cosh potential they used 
  has an unphysical long tail as seen in Fig.~4 
(the volume integral is 695 MeVfm$^3$!)
and   is not appropriate for the description of the energy levels of the 
highly excited  cluster states  and 
 $^3$He+$^{16}$O scattering.  
 In fact, their calculated energy levels are located
 at  energies which are lower than they should be if an appropriate potential is used.
  (The drawback of this long tail of the cosh
 potential has already been discussed  
   for  the $\alpha$+$^{40}$Ca system \cite{Ohkubo1989} .)
Although they tried to interpret their observed states as  member states of the
 $N=9$ band unsuccessfully,
the observed energy levels should be regarded as the $N=8$ higher nodal states
 with $L=$2, 4 and 6. The correspondence between the observed states and 
 the present calculation is good as seen  
in Fig.~3.
The observed states at $E_x$=13.8, 14.88 and 16.24 MeV  have 
comparable widths
$\Gamma_{c.m.}$=0.67, 0.62 and 0.40 MeV, respectively and  have a large   $L=2$
 contribution
 in the Legendre polynomial analysis of the angular distributions of  the 
$^{16}$O($^3$He,$\gamma$) capture cross sections \cite{Waltham1983}. These states
 may be  considered to be
fragmented from the higher nodal $L=2$ state. (The centroid energy is
15 MeV.) The state at $E_x$=18.4 MeV with $\Gamma_{c.m.}$=4.4 MeV  
 \cite{Waltham1983} has a  broad resonant structure and 
may be considered as  a candidate for the 
 member state of the higher nodal band with  $L=4$. 
The dimensionless reduced width $\theta^2$ 
(defined by $\Gamma=2P_L(a)\gamma_W^2(a)\theta^2(a)$, with $P$ the Coulomb penetrability, 
 $\gamma_W^2$ the Wigner limit value and $a$ the channel radius) calculated at a
 channel radius 5 fm 
is  $\theta^2$ =0.20, 0.13 and 0.07 for the $E_x$=13.8, 14.88 and 16.24 MeV states respectively,
 assuming $L=2$ and  $\theta^2$=1.0 for the 
$E_x$=18.4 MeV state assuming $L=4$, which is compatible with the
present picture. 
 The rotational constant of the calculated $N=8$ band,  $k\simeq$0.21, 
  is comparable with  $k\simeq$0.24 estimated from the experimental 
  centroid at $E_x$=15 MeV  ($L=2$) and  the  
  18.4 MeV ($L=4$) state.
 Yamazaki {\it et al.} \cite{Yamazaki2007} claim  that they observed  
 six prominent peaks 
 above $E_x$= 12 MeV in the $^{16}$O($^6$Li,t) transfer reactions.
 It is interesting to know  whether these   could be  fragmented 
  states of the $N=8$ higher nodal band.

To summarize, 
 the existence of the $N=8$  higher nodal band states with the 
 $^3$He+$^{16}$O cluster 
structure in $^{19}$Ne, in which inter-cluster  relative motion is excited,  
was strongly suggested  by studying  $^3$He+$^{16}$O scattering 
which shows prerainbow oscillations.
 The calculated low  spin members of the higher nodal band states
 with $L=0$ correspond well with the observed state in the low energy $^3$He elastic
 scattering at $E_L$=5.05 MeV  \cite{Roepke1967,Ajzenberg-Selove1987}
 and the states with 
 $L=2$, 4 and 6 correspond well to the observed states in  the 
 $^{16}$O($^3$He,$\gamma$)  capture  reactions  \cite{Waltham1983}.
 A higher nodal band state may 
appear more clearly and stably  in $^{19}$Ne than  in $^{20}$Ne due to coupling
 to the other  states nearby.
 In the mirror nucleus  $^{19}$F  a similar  higher nodal
 band with the t+$^{16}$O cluster structure and  prerainbow oscillations 
in t+$^{16}$O scattering are expected.
The present findings   about the   higher nodal band states  with the   $^3$He 
cluster   in $^{19}$Ne in 
addition to the higher nodal states with the  $\alpha$ cluster structure in the 
 $^{20}$Ne, $^{40}$Ca  and $^{44}$Ti nuclei  and the higher nodal states  
with the $^{16}$O cluster structure 
in the $^{32}$S nucleus   reinforce the importance of the concept of the higher nodal
state and the  $^3$He cluster  in  nuclei.

\par
   One of the authors (S.O.)   has been supported by a
 Grant-in-aid for Scientific Research
 of the Japan Society for Promotion of Science (No. 16540265)
 and the Yukawa Institute for Theoretical Physics .

\end{document}